# Topic Classification Method for Analyzing Effect of eWOM on Consumer Game Sales

Yoshiki Horii
Department of Information and
Management System Engineering
Nagaoka University of Technology
Nagaoka, Japan
s153417@stn.nagaokaut.ac.jp

Hirofumi Nonaka
Department of Information and
Management System Engineering
Nagaoka University of Technology
Nagaoka, Japan
nonaka@kjs.nagaokaut.ac.jp

Elisa Claire Alemán Carreón
Department of Information and
Management System Engineering
Nagaoka University of Technology
Nagaoka, Japan
s153400@stn.nagaokaut.ac.jp

Hiroki Horino
Department of Information and
Management System Engineering
Nagaoka University of Technology
Nagaoka, Japan
s153418@stn.nagaokaut.ac.jp

Toru Hiraoka
Department of Information Systems
University of Nagasaki
Nagasaki, Japan
hiraoka@sun.ac.jp

*Abstract*—Electronic word-of-mouth (eWOM) has become an important resource for analysis of marketing research. In this study, in order to analyze user needs for consumer game software, we focus on tweet data. And we proposed topic extraction method using entropy based feature selection based feature expansion. We also applied it to the classification of the data extracted from tweet data by using SVM. As a result we achieved a 0.63 F-measure.

## I. INTRODUCTION

With the rapid growth of Internet technology, electronic word-of-mouth (eWOM), which means consumers opinion through web service related to the goods and services, have become an important resource for analysis of marketing research. There are many previous studies on eWOM in many product and service categories. These researches indicate that eWOM does not only influence on buying behavior of consumers, but also on commercial results. Pelsmacker, et.al [1], analyze reviews of hotel industry and find that review volume drives room occupancy and review valence impacts RevPar and Digital marketing strategies.

In recent years, Text mining is widely used for economy and management researches. Bollen et al., [2] predicted the Dow Jones Industrial Average by extracting public mood from tweets. They used the mood tracking tool GPOMS and achieved high accuracy in predicting the daily changes in the closing values of the Dow Jones Industrial Average. There are several other studies on relationship between text data and economic data. Asur et al., [3] forecasted box-office revenues for movies using Twitter data. Mao [4] used Mobile communication data to measure socio-economic indicators in. O'Connor et al., [5] conducted correlation analysis between consumer confidence and sentiment word frequencies in contemporaneous tweets.

Nonaka et al, [6] developed a patent scoring method based on citation networks and analyzed relation between the score and stock price.

On the other hand, some researchers attempted to analyze user needs by using Text mining. Hea et.al., [7] analyzed web marketing of pizza industry by text-mining tools applying to Tweets and Facebook data. Nonaka et al [8, 9] extracted technology words and user needs from patent documents by using an entropy and grammatical pattern based method.

While Text mining approach is widely used for economy and management researches, there are few researches using text data in this field. Zamani et al., [10] attempted to predict real estate at the level of US country using the LDA [11] topics of tweets. However, because tweets are not only related to real estates, this research did not focus on eWOM.

In this study, in order to analyze user needs for consumer game software, we focus on tweet data. And we proposed topic extraction method using entropy based feature selection based feature expansion. We also applied it to the classification of the data extracted from tweet data by using SVM.

This paper is organized as follows. In section 2, we explain about methodology. We present our experiments of system performance and analysis result in section 3. Based on this system, we conduct an evaluation including the discussions in section 4. Finally, section 5 concludes this paper.

## II. METHODOROGY

### 2.1 Word Segmentation

For an analysis to be made possible for each word, we segmented the collected Japanese texts without spaces into words using a Japanese morphological analyzer tool called MeCab [12]. After segmenting the words, we extracted only self- sufficient words.

### 2.2 Entropy Based Keyword Extraction

Feature selection of our method is based on the Shannon's entropy (hereinafter referred as entropy) value [13] of each word. According to information theory, entropy is the expected value of the information content in a signal.

Applying this knowledge to the study of words allows us to observe the probability distribution of any given word inside the corpus. For example, a word that keeps reappearing in many different documents will have a high entropy, while a word that only was used in a single text and not in any other documents in the corpus will bear an entropy of zero. This concept is shown in Figure 2.

Having previously tagged a sample of texts positive and negative by pertinence to each category, if a word has higher entropy in positive documents than in negative documents by a factor of alpha greater than 1 (α > 1), then it means its probability distribution is more spread in positive texts, meaning that it is commonly used in positive tagged documents compared to negative ones.

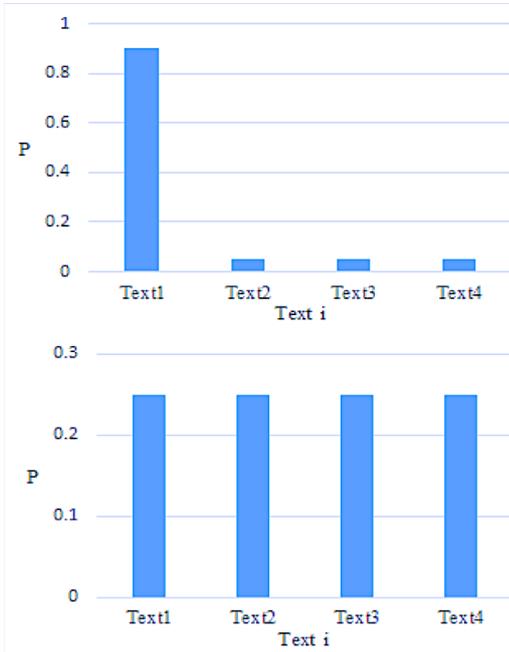

**Figure 2: Probabilities of a word j being contained in a document i. Up. Entropy close to zero, Down. High entropy**

To calculate the entropy in a set of documents, for each word j that appears in each document i, we counted the number of times a word appears in positive comments as $N_{ijP}$, and the number of times a word appears in negative comments as $N_{ijN}$. Then, as shown in the formulas below, we calculated the probability of each word appearing in each document shown below as $P_{ijP}$ (1) and $P_{ijN}$ (2).

$$P_{ijP} = \frac{N_{ijP}}{\sum_i^M N_{ijP}} \quad (1)$$

$$P_{ijN} = \frac{N_{ijN}}{\sum_i^M N_{ijN}} \quad (2)$$

We then substitute these values in the formula that defines Shannon's Entropy. We calculated the entropy for each word j in relation to positive documents as $H_{Pj}$ (3), and the entropy for each word j in relation to negative texts as $H_{Nj}$ (4). That is, all instances of the summation when the probabilities $P_{ijP}$ or $P_{ijN}$ are zero and the logarithm of these becomes undefined are substituted as zero into (3) and (4).

$$H_{Pj} = -\sum_{i=1}^{M}[P_{ijP} log_2(P_{ijP})] \quad (3)$$

$$H_{Nj} = -\sum_{i=1}^{M}[P_{ijN} log_2(P_{ijN})] \quad (4)$$

After calculating the positive and negative entropies for each word, we measured their proportion using the mutually independent coefficients α for positive keywords and α' for negative keywords, for which we applied several values experimentally. A positive keyword is determined when (5) is true.

$$H_{pj} > \alpha H_{Nj} \quad (5)$$

### 2.3 Word Analysis Using Support Vector Machine

In machine learning, Support Vector Machines are supervised learning models commonly used for statistical classification or regression [14]. Using already classified and labeled data with certain features and characteristics, an SVM learns to classify new unlabeled data by drawing the separating (p-1)-dimensional hyperplane in a p-dimensional space. Each dimensional plane is represented by one of the features that a data point holds. Then each data point holds a position in this multi-dimensional space depending on its features. The separating hyperplane and the supporting vectors divide the multi-dimensional space by minimizing the error of classification. A two-dimensional example is shown below.

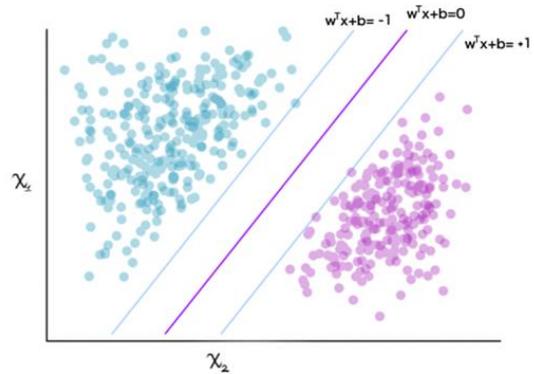

**Fig 3.** 2D example of the Linear SVM classification

The linear kernel for the SVM classification is defined by the formula (6) below. The influence that each point of the training data inputs into the vector is defined by their weight $w_n$, included in the Weight Vector $w$. The bias coefficient b determines the position of the hyperplane.

$$f(x) = w^T x + b \quad (6)$$

Then, the conditions shown in (7) are applied when classifying new data.

$$f(x) \begin{cases} \geq 0 & y_i = +1 \\ < 0 & y_i = -1 \end{cases} \quad (7)$$

Now, the initial condition is set as w = 0. Then each possible separating vector is tested, and when a classification for $x_i$ fails, the value for w is changed as follows in (8) by a value of α.

$$w \leftarrow w + \alpha sign(f(x_i))x_i \quad (8)$$

## III. RESULTS

### 3.1 Experiments of system performance

Before analysis by using Support Vector Machines, we defined five topics by manually. In Twitter posts are mainly mentioned about five topics. We made the training data by randomly collecting 400 posts and classified into five topics.

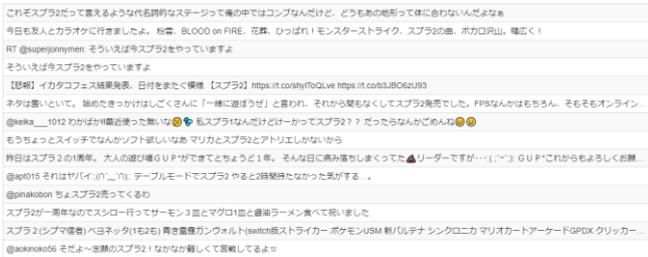

**Fig 4.** Example posts in Twitter

Table 1 shows the topic content and example keywords for "About impressions" topic obtained from Entropy.

Table 1
Topic content and example keywords of each topic

|  | Topic content | Example keywords |
|---|---|---|
| Topic 1 | Included URL | Include "http://", "https://",".com" in acquired tweet expanded URL |
| Topic 2 | Include photo | Include "photo" in acquired tweet entities status |
| Topic 3 | About impressions | "面白い"(*omoshiroi*, interesting) "つまらない"(*tsumaranai*, boring) |
| Topic 4 | Retweeted | Include "RT" in acquired tweet first text |
| Topic5 | Reply | Include "@" in acquired tweet first text |

We crawled a total of 487,602 posts from Twitter. Twitter has a high affinity with consumer games, and these topics are frequently posted in Japan.

We classified posts into negative or positive about impressions topics shown on above by using Support Vector Machines. To evaluate our trained machines, we calculated then the Precision, Recall, $F_1$-Score. [15,16]

Precision($P$) is defined as the number of true positives ($T_p$) over the number of true positives plus the number of false positives ($F_p$).

$$Precision(P) = \frac{T_p}{T_p + F_p}$$

Recall($R$) is defined as the number of true positives ($T_p$) over the number of true positives plus the number of false negatives ($F_n$).

$$Recall(R) = \frac{T_p}{T_p + F_N}$$

$F_1$-Score is a measure of a test's accuracy, and it considers both the Precision and the Recall of the test to compute the score. $F_1$-Score is the harmonic mean of Precision and Recall, and it is can be calculated from the following formula.

$$F_1 - Score = \frac{2 * Precision * Recall}{Precision + Recall}$$

Table 2 shows the results of the Precision, Recall, $F_1$-Score in "About impression" topic.

Table 2
Results of the Precision, Recall, $F_1$-Score in "About impressions" topic

|  | Precision | Recall | $F_1$-Score |
|---|---|---|---|
| About impressions | 0.46 | 1.00 | 0.63 |

### 3.2 Result of Analysis

We crawled a total of 487,602 posts from Twitter for data analysis of user needs. And then we applied SVM to classify "About impressions" topics. Tweets categorized as negative were very few compared to tweets classified as positive.

## IV. DISCUSSION

First, we focus on the evaluation of system performance. Our method achieved 0.64 of the $F_1$-Score of "About impressions" topics and 0.46 of the Precision which can be considered as high. Then we think our method can be used for real data analysis. However, there is still room for improvement of Precision. This result means that although data was not missed, more than half of the data was erroneously judged. Therefore, in order to solve the issue, we will have to increase training data.

## V. CONCLUSION AND FUTURE WORKS

In our study, we classified tweets collected as a basic research to analyze the relevance between the sales of consumer games and the posted contents of Twitter. Precision was low, but the Recall rate was 100 percent. Improvement of precision can be expected by increasing training data.

For our future work, we need to improve the Precision and $F_1$-Score. To achieve this, we need to increase training data. There is also room for improvement in the way of classification. In addition, we would like to add movie and playing status topics.